\documentstyle[aps,amsfonts,pre,multicol]{revtex}
\tighten
\draft
\newtheorem{Def}{Definition}
\newtheorem{Rul}{Rule}
\newtheorem{Lemm}{Lemma}

\newtheorem{Thrm}{Theorem}
\begin{document}
\title{Finding The Sign Of A Function Value By Binary Cellular Automaton}
\author{H. F. Chau\footnote{Corresponding author. Email address:
 hfchau@hkusua.hku.hk}, Hao Xu, K. M. Lee, L. W. Siu and K. K. Yan}
\address{Department of Physics, University of Hong Kong, Pokfulam Road, Hong
 Kong}
\date{\today}
\maketitle
\begin{abstract}
 Given a continuous function $f(x)$, suppose that the sign of $f$ only
 has finitely many discontinuous points in the interval $[0,1]$. We show how to
 use a sequence of one dimensional deterministic binary cellular automata to
 determine the sign of $f(\rho)$ where $\rho$ is the (number) density of $1$s
 in an arbitrarily given bit string of finite length provided that $f$
 satisfies certain technical conditions.
\end{abstract}
\medskip
\begin{multicols}{2}

\section{Introduction}
\label{S:Intro}
 Cellular Automaton (CA) is a simple local interaction model of natural systems
 used extensively in various fields of physics\cite{Wolfram}. Besides, CA can
 be regarded as a computation model motivated by biological phenomenon. Various
 inequivalent definitions of CA exist in the community. In this paper, we
 restrict ourselves to consider the time evolution of CA to be governed by a
 local synchronous deterministic uniform rule. In other words, the state of
 each site in the next time step depends deterministically only on the states
 of its finite neighborhood, the states of all sites are updated in parallel,
 and the transformation CA rule table is covariant under translation of the
 background lattice. In this respect, CA can be regarded as a decentralized
 deterministic parallel computation model without central memory storage. It is
 useful to study its power and limitation. In fact, a
 recent renewal interest in CA computing\cite{Comp} makes such an
 investigation timely.

 Since CA rules are local in nature, it is instructive to see if CA can perform
 tasks that involve global quantities. Perhaps the most well-known example is
 the so-called density classification problem. Considering a
 bit string of finite length $N$ with periodic boundary conditions, the problem
 is to change all bits in the string to $0$ if the number of $0$s is greater
 than the number of $1$s in the input bit string, and to change all bits to $1$
 if the number of $1$s is greater than the number of $0$s. Clearly, the global
 quantity involved in this problem is the density of $1$s which is defined as
 the number of $1$s in the string divided by the string length $N$.

 Various CA rules have been proposed to solve the density classification
 problem both by human\cite{GKL} and by genetic algorithm\cite{Mitchell}. But
 they only provide approximate solutions. In other words, these rules work for
 most but not all randomly chosen initial configurations. In fact, Land and
 Belew showed that density classification cannot be performed perfectly using a
 single one dimensional CA rule\cite{LandBelew}. Later on, Capcarrere and his
 collaborators proved that a single CA rule can solve the density
 classification and other related problems exactly if we modify either the
 required output of the automaton or the boundary
 conditions\cite{Capcarrere,Sipper}. However, in their
 approach, it has to scan through the states of all sites in the final
 configuration, in general, before knowing the answer. This requires global
 memories in the read out process. In contrast, read out in
 the original density classification problem can clearly be done by looking at
 the states of a few local sites, and hence can be done with an additional
 finite rule table whose size is independent of $N$.

 Recently, Fuk\'{s} pointed out that the density classification problem can be
 solved exactly if we apply two CA rules in succession\cite{Fuks}. More
 precisely, he showed that by applying a CA rule a fixed number of times
 depending only on the lattice size and then followed by applying another CA
 rule a fixed number of times depending again only on the lattice size, the
 density classification problem can be solved exactly. Fuk\'{s}
 further asked if it is possible to classify a rational density. That is, he
 questioned if it is possible to determine,
 using succession of CA rules, whether
 the density of $1$s in an arbitrary one dimensional array with periodic
 boundary conditions is less than, equal to or greater than a prescribed
 rational number $p/q$, called the critical density. Chau {\it et al.} answered
 his question affirmatively by showing that two CA rules are necessary and
 sufficient in solving the rational density classification problem
 exactly\cite{RationalDC}. His group further generalized their algorithm to
 find the majority state of an one dimensional array when each site may take
 on $n$ possible states \cite{naryDC}. In addition, further constraints on the
 forms of certain CA rules that can be used in the rational density
 classification problem have also be reported\cite{Boccara}.

 In Section~\ref{S:Problem}, we shall spell out the details of the problem that
 we are interested in. Then after introducing some technical results in
 Section~\ref{S:BuildingBlock_CAs}, we shall solve the problem using a set of
 binary CAs with local synchronous deterministic uniform rules in
 Section~\ref{S:Solution}. The general solution reported in
 Section~\ref{S:Solution} can be simplified in a number of special but
 important cases. We report the efficient solutions to these special cases in
 Section~\ref{S:Generalization}. Finally, we give a brief discussions on the
 significance of our result in Section~\ref{S:Discussion}.

\section{Statement Of The Problem}
\label{S:Problem}
 In this paper, we propose another generalization of the density classification
 problem, called the generalized density classification problem (GDCP) and solve
 it using a sequence of CA rules. Before we discuss the significance of our
 generalization, let us first state our problem precisely.

 Suppose we are given an input bit string $\alpha$ of length $N$ in periodic
 boundary conditions, as well as a continuous function $f(x)$ with finitely
 many discontinuous points of $\mbox{sgn} (f(x))$ in the interval $[0,1]$,
 where
\begin{equation}
 \mbox{sgn} (y) = \left\{ \begin{array}{rl}
  1 & \mbox{if~} y > 0, \\
  -1 & \mbox{if~} y < 0, \\
  0 & \mbox{if~} y = 0.
 \end{array} \right. \label{E:sgn_def}
\end{equation}
 Moreover, those discontinuous points are rational and shall be denoted by
 $p_i/q$ where $p_i,q\in {\Bbb Z}^{+} \cup \{ 0 \}$, and $i$ is the index for
 the discontinuous points. In addition, we have the freedom to choose the
 smallest positive $q$ for the denominators of all the expressions $p_i/q$.
 With this choice, $p_i$ and $q$ need not be relatively prime. Finally, we
 require that $f(\frac{1}{2q})$ and $f(\frac{2q-1}{2q})$ are non-zero.

 We define the density of $1$s of the bit string $\alpha$ as the total number
 of $1$s in the bit string divided by the string length $N$ and is denoted by
 the symbol $\rho (\alpha)$. We may simply write $\rho$ instead of $\rho
 (\alpha)$ when it is clear from the content which bit string we are referring
 to.

 The task of GDCP is to evolve the bit string $\alpha$ to all $1$ (or $0$) if
 $\mbox{sgn} (f(\rho (\alpha))) = 1$ (or $\mbox{sgn} (f(\rho (\alpha))) = -1$).
 Besides, we have to preserve the density of $1$s of the bit string if
 $\mbox{sgn} (f(\rho (\alpha))) = 0$. In addition, by inspecting a fixed
 portion of the final output string, we are able to distinguish between the
 above three cases with certainty. (We shall consider a slightly relaxed
 variant to this definition later on in Section~\ref{S:Generalization}.) To
 solve the GDCP by a sequence of binary CA rules, we mean that every initial
 configuration can be correctly classified by applying some binary CA rule
 $R_1$ $t_1$ times followed by another binary CA rule $R_2$ $t_2$ times and so
 on up to the $k$th binary CA rule $R_k$ $t_k$ times, where $k$ is fixed and
 independent of the string length $N$. Besides, each CA rule $R_i$ must be
 independent of the input $\alpha$ and has a bounded rule table size that is
 also independent of $N$. Finally, the number of iterations $t_i$ is
 independent of the initial configuration, but may depend on $N$.

 Clearly, the GDCP reduced to the rational density classification problem by
 choosing $f(x) = x - \rho_c$ where $\rho_c$ is a rational number. Moreover,
 the GDCP can be solved trivially if we have a global counter.

 In addition to naturally generalize the density classification problem, GDCP
 is interesting on its own right. First, by choosing
\begin{equation}
 f(x) = \left\{ \begin{array}{cl}
  x-\rho_{c1} & \mbox{if~} x < \rho_{c1}, \\
  0 & \mbox{if~} \rho_{c1} \leq x \leq \rho_{c2}, \\
  x-\rho_{c2} & \mbox{if~} x > \rho_{c2},
 \end{array} \right. \label{E:CoarseGrainDC}
\end{equation}
 where $\rho_{c1} < \rho_{c2}$ both being rational, it determines if the
 density of $1$s is in the interval $(\rho_{c1},\rho_{c2})$ from a single copy
 of input bit string. (Note that it is more efficient to ask if $\rho$ is
 between $1/10$ and $4/39$ rather than to ask if $\rho$ equals $0.10215$ partly
 because of our result in Section~\ref{S:Solution} that the former question
 requires smaller CA rule tables.) Thus, GDCP can be used to determine the
 coarse-grained density of $1$s efficiently, while in the rational density
 classification problem, the resolution is infinite in the sense that we can
 only determine whether $\rho>\rho_c$, $\rho=\rho_c$ or $\rho<\rho_c$, no
 matter how closed $\rho$ is to $\rho_c$. This reason alone is good enough to
 investigate the GDCP. Second, the GDCP has a more interesting flow diagram.
 While the rational density classification automata has two stable and one
 unstable fixed points in the density of 1s, the GDCP automata in general has
 at most two stable fixed points ($\rho=1$ and $\rho=0$), finitely many
 unstable fixed points (at where $f(\rho)=0$ but $f$ is not identical zero
 locally near $\rho$) and infinitely many neutral fixed points (at where
 $f(\rho)=0$ and $f$ is identical zero locally near $\rho$) in the limit of
 $N\rightarrow \infty$.

\section{Some Useful Cellular Automaton Building Blocks}
\label{S:BuildingBlock_CAs}
 Let us first report several essential building blocks of our solution to the
 GDCP.

\subsection{Car Hopping Automaton}
\label{SS:Traffic}
 The first step in solving the problem
 is to make the density of 1s uniform so as
 to make the local density of 1s a good estimation of the global density of 1s
 in the bit string. To measure the local density, we introduce the concept of
 $k$th order local number. It is just the number of 1s around the site we are
 interested. The CA rule ${\mathbf H}_k$ will move the 1s to the right
 according to the local density gradient. If we repeatedly apply
 ${\mathbf H}_k$ to an arbitrary bit string, eventually the density of 1s of
 the resulting string will be more or less uniform. Of course we cannot expect
 that all local numbers are equal, but we can prove that the difference between
 local numbers of different sites can only be less than or equal to one. This
 CA rule is the most important CA rule in this paper.

\begin{Def}
 We denote a configuration in our bit string of length $N$, which can be
 identified with an one dimensional array of sites in periodic boundary
 conditions, by $\alpha$ and the state of the $i$th site by $\alpha(i)$. That
 is, $\alpha(i)$ is the value of the $i$th bit in the string counting from left
 to right. We define the $k$th order local number of the $i$th site in our one
 dimensional array by $n_k (\alpha,i) = \sum_{j=0}^{k-1} \alpha(i+j)$ where the
 sum $i+j$ in the index here and elsewhere in this paper is understood to be
 modulo $N$.

 In other words, the $k$th order local number is the number of sites with
 state $1$ among the $k$ consecutive sites starting from the site $i$. So, if
 we regard state $0$ as an empty site and state $1$ as a site with a car, the
 $k$th order local number gives the local number of cars near a site. From
 now on, we shall interchangeably describe the state $1$ as a car and the state
 $0$ as an empty site. Thus, the density of $1$s of a bit string can also be
 referred to as the car density.

 In what follows, we shall write $n_k (i)$ instead of $n_k (\alpha,i)$ when
 there is no confusion on which bit string we are referring to. 
 \label{Def:LocalNumber}
\end{Def}
\begin{Rul}[$k$th order car hopping rule]
 Let us denote this CA rule by ${\mathbf H}_k$. The state of site $i$ in the
 next time step ${\mathbf H}_k [\alpha](i)$ is given by:
 \begin{eqnarray}
  {\mathbf H}_k [\alpha](i) = \left\{ \begin{array}{cl}
   0 & \mbox{if~} \alpha(i) = 1, \alpha(i+1) = 0 \mbox{~and~} \\
   & ~ n_k(i-k+1) > n_k(i+1), \\
   1 & \mbox{if~} \alpha(i) = 0, \alpha(i-1) = 1 \mbox{~and~} \\
   & ~ n_k(i-k) > n_k(i), \\
   \alpha(i) & \mbox{otherwise}.
  \end{array} \right. \label{E:TrafficRule_CA}
 \end{eqnarray}
 This CA rule can be rewritten in another form as follows: we
 interchange the states in site $i$ and $i+1$ if and only if $\alpha(i) = 1,
 \alpha(i+1) = 0$ and $n_k (i-k+1) > n_k(i+1)$. All other states remain
 unchanged. All updates are taken in parallel. In what follows, we shall
 sometimes refer to such an interchange as hopping of a car.
 \label{Rul:Traffic}
\end{Rul}

 For example, ${\mathbf H}_4 (011001010010) = 010100101010$.  

 Note that the role of $0$ and $1$ in the above CA rule is symmetric in the
 sense that it is invariant under the interchange of state $0$ and $1$ plus the
 relabeling of $i$ as $-i$. Following the spirit of particle physics, we shall
 refer to this property as the CP symmetry. Furthermore, the car density $\rho$
 is conserved under this CA rule. In fact one way to interpret this CA rule is
 that cars are driven to move from left to right by local car density gradient.
 Or equivalently, empty sites are driven to move from right to left by local
 empty site density gradient. Therefore, we expect that the $k$th order local
 numbers in all recurrent states under the repeated iteration of
 ${\mathbf H}_k$ is evenly distributed. Nonetheless, a rigorous proof turns out
 to be rather involved partly because a car at site $i$ cannot hop whenever
 another car is occupying site $i+1$ even in the presence of a car density
 gradient. We now begin rigorous proof by first introducing a few technical
 lemmas.

 We expect that for any bit string $\alpha$, the density of $\beta \equiv
 {\mathbf H}_k^l (\alpha)$ will be uniform if $l$ is large enough. The first
 lemma shows that this is indeed the case when $\beta$ is a fixed point of
 ${\mathbf H}_k$.
\begin{Lemm}
 The $k$th order local number $n_k (i)$ is equal for all $i$ if and only if
 ${\mathbf H}_k (\alpha) = \alpha$ (that is, no car hops under the action of
 the $k$th order car hopping rule). \label{Lemm:UniformDensity}
\end{Lemm}
\noindent
{\it Proof:} If ${\mathbf H}_k (\alpha) \neq \alpha$, then some car in the bit
 string $\alpha$ hops under the action of ${\mathbf H}_k$. So clearly $n_k (i)
 > n_k (i+k)$ for some $i$.

 Conversely, if ${\mathbf H}_k (\alpha) = \alpha$, then we may assume that
 $\alpha$ does not equal to all $0$ or all $1$, for otherwise the Lemma is
 trivially true. Now locally we have any one of the following three situations.

 Case~(a) $\alpha (i-1) = 1$ and $\alpha (i) = 0$: in this case we clearly
 require $n_k (i-k) \leq n_k (i)$ in order to prevent the car located at site
 $i-1$ from hopping.

 Case~(b) $\alpha (i) = \cdots = \alpha (i+j-1) = 1$ and $\alpha (i+j) = 0$
 for some $j > 0$: in this case we again conclude that $n_k (i-k) \leq n_k (i)$
 for otherwise we have $n_k (i+j-k) \geq n_k (i+j-k-1) \geq \cdots \geq n_k
 (i-k) > n_k (i) \geq n_k (i+1) \geq \cdots \geq n_k (i+j)$ contradicting
 the assumption that the car located at site $i+j-1$ does not move.

 Case~(c) $\alpha (i) = \alpha (i-1) = \cdots = \alpha (i-j) = 0$ and $\alpha
 (i-j-1) = 1$ for some $j > 0$: the CP symmetry of the CA rule ${\mathbf H}_k$
 implies that we can use the same trick as in the proof of case~(b) to show
 that $n_k (i-k) \leq n_k (i)$.

 In summary, we conclude that $n_k (i-k) \leq n_k (i)$ for all $i$; and since
 we are working in a finite bit string with periodic boundary, this is
 possible only when $n_k (i-k) = n_k (i)$ for all $i$.

 Suppose that $n_k (i-k+1) > n_k (i-k+2)$ for some $i$, then $\alpha (i-k+1) =
 1$ and $\alpha (i+1) = 0$. However, this also implies that $n_k (i+1) >
 n_k (i+2)$ and hence $\alpha (i+1) = 1$ and $\alpha (i+k+1) = 0$ which is a
 contradiction. By the same token, it is absurd to have $n_k (i-k+1) < n_k
 (i-k+2)$. Thus, the only possibility is that $n_k (i) = n_k (i+1)$ for all $i$
 and hence $n_k (i) = n_k (j)$ for all $i,j$.
\hfill $\Box$

\bigskip
 Lemma~\ref{Lemm:UniformDensity} gives an important property of fixed points of
 ${\mathbf H}_k$. To characterize the property of cycles of ${\mathbf H}_k$
 with period greater than one, we need to work harder. In what follows, we show
 that whenever there is a fluctuation in the $k$th order local number $n_k (i)$
 in a bit string string $\alpha$, at least one car in this string must hop
 under the action of the $k$th order car hopping rule ${\mathbf H}_k$.
\begin{Lemm}
 If $n_k (\alpha,i-k+1) > n_k (\alpha,i+1)$, then there exists $j$ with $i-k+1
 \leq j < i+k$ with $n_k (\alpha,j-k+1) \geq n_k (\alpha,i-k+1)$ such that
 there is a car located at site $j$ and this car hops by applying the rule
 ${\mathbf H}_k$. Moreover, a car will hop from site $i$ to $i+1$ in no more
 than $k$ time steps. That is to say, there exists $r\leq k$ such that
 ${\mathbf H}_k^{r-1} [\alpha] (i) = 1$, ${\mathbf H}_k^{r-1} [\alpha] (i+1) =
 0$ and ${\mathbf H}_k^r [\alpha] (i) = 0$. \label{Lemm:Exist}
\end{Lemm}
\noindent
{\it Proof:} We divide the proof into the following three cases.

 Case~(a) $\alpha (i) = 1$ and $\alpha (i+1) = 0$: we choose $j = i$ as the
 car located at site $i$ hops by applying ${\mathbf H}_k$.

 Case~(b) $\alpha (i+1) = 1$: since $n_k (i-k+1) >  n_k (i+1)$, we can always
 find $1 \leq m < k$ such that $\alpha (i+1) = \alpha (i+2) = \cdots = \alpha
 (i+m) = 1$ and $\alpha (i+m+1) = 0$. Then it is easy to see that $n_k
 (i+m-k+1) \geq n_k (i-k+1) > n_k (i+1) \geq n_k (i+m+1)$. So the car located
 at site $i+m$ hops to $i+m+1$ when we apply ${\mathbf H}_k$. Hence, we choose
 $j = i+m$. Inductively, in the next time step, we may choose $j = i+m-1$ and
 so on. Consequently, in no more than $k$ time steps, a car must hop from site
 $i$ to $i+1$.

 Case~(c) $\alpha (i) = \alpha (i+1) = 0$: exploiting the CP symmetry of the
 $k$th order car hopping rule, this case can be proved in the same way as in
 case~(b).
\hfill$\Box$

\bigskip
 We now introduce the fluctuation amplitude, which is the difference of the
 local numbers in different sites. It will be shown that the maximum
 fluctuation amplitude will not increase by the repeated application of
 ${\mathbf H}_k$.
\begin{Def}
 Let $\alpha$ be a bit string of length $N$. We denote $\Delta_k (\alpha) =
 \max \{ n_k (\alpha,i) - n_k (\alpha,j) : 1\leq i,j\leq N \}$ the maximum
 fluctuation amplitude of the $k$th order local number. We write the maximum
 and minimum $k$th order local numbers of a bit string by $n_k^{\rm (max)}
 (\alpha) = \max \{ n_k (\alpha,i) : 1\leq i\leq N\}$ and $n_k^{\rm (min)}
 (\alpha) = \min \{ n_k (\alpha,i) : 1\leq i\leq N\}$, respectively. In
 addition, we denote the number of sites with $k$th order local number equals
 $m$ by $M_k (\alpha,m) = |\{ 1\leq i\leq N : n_k (\alpha,i) = m \}|$.

 For simplicity, we shall drop the label $\alpha$ in $\Delta_k$,
 $n_k^{\rm (max)}$, $n_k^{\rm (min)}$ and $M_k$ when it is clear which bit
 string we are referring to in the text. \label{Def:Fluctuation}
\end{Def}

\bigskip
 Now, we are ready to show that repeated application of ${\mathbf H}_k$
 decreases the fluctuation amplitude of the $k$th order local number to its
 minimum possible value in $\mbox{O} (N)$ time steps. More precisely, we are
 going to prove the following lemma:
\begin{Lemm}
 Let $\alpha$ be a bit string of length $N$. Then $\Delta_k
 ({\mathbf H}_k^{\ell (k)}(\alpha)) \leq 1$
 where $\ell \equiv \ell (k) \equiv 2k (k-2)
 \left\lceil \frac{1}{2} \left\lceil \frac{N}{k} \right\rceil \,\right\rceil$.
 Besides, $\Delta_k ({\mathbf H}_k^{\ell (k)}(\alpha)) = 0$
 if and only if $\rho (\alpha)
 = r / k$ for some $r\in {\Bbb Z}$. \label{Lemm:Fluctuation}
\end{Lemm}
\noindent
{\it Proof:} First of all, we show that if $\Delta_k (\beta) \leq 1$, then
 $\Delta_k (\beta) = 0$ if and only if $\rho (\beta) = r/k$ for some $r\in
 {\Bbb Z}$. Since the fluctuation of $n_k (\beta,i)$ is less than 2, $n_k
 (\beta,i)$ can either be $\left\lfloor k\rho (\beta) \right\rfloor$ or
 $\left\lceil k\rho (\beta) \right\rceil$. So, if $\rho (\beta) = r/k$ for some
 $r\in {\Bbb Z}$, then $\Delta_k (\beta) = 0$. Conversely, if $\rho (\beta)$ is
 not in the form $r/k$ for some $r\in {\Bbb Z}$, then there exist $i$ and $j$
 such that $n_k (\beta,i) = \left\lfloor k\rho (\beta) \right\rfloor$ and $n_k
 (\beta,j) = \left\lceil k\rho (\beta) \right\rceil$. Hence, $\Delta_k (\beta)
 = 1$.

 Since ${\mathbf H}_k$ conserves car density, the last statement of this Lemma
 is proved by setting $\beta = {\mathbf H}_k^\ell (\alpha)$.

 To prove the first part of this Lemma, we may further assume that
 (a)~${\mathbf H}_k^i (\alpha) \neq {\mathbf H}_k^{i-1} (\alpha)$ for all $i <
 \ell$ and (b)~$\Delta_k (\alpha) > 1$. For if~(a) does not hold, then
 ${\mathbf H}_k^\ell (\alpha)$ is a fixed point of ${\mathbf H}_k$ and hence
 Lemma~\ref{Lemm:UniformDensity} tells us that $\Delta_k ({\mathbf H}_k^\ell
 (\alpha)) = 0$. In addition, Eq.~(\ref{E:TrafficRule_CA}) tells us that it is
 not possible to increase the $k$th order local number $n_k (\beta,i)$ of a
 site $i$ in a bit string $\beta$ if $n_k (\beta,i)$ equals $n_k^{\rm (max)}
 (\beta)$. Hence, $n_k^{\rm (max)} ({\mathbf H}_k^i (\alpha))$
 is a non-increasing
 function of $i$; and by CP symmetry of the $k$th order car hopping rule,
 $n_k^{\rm (min)} ({\mathbf H}_k^i (\alpha))$
 is a non-decreasing function of $i$.
 As a result, the first part of this Lemma is trivially true if $\Delta_k
 (\alpha) \leq 1$.

 Eq.~(\ref{E:TrafficRule_CA}) also tells us that for any bit string $\beta$,
 $n_k ({\mathbf H}_k (\beta),i) = n_k (\beta,i)+1$ if and only if a car in
 $\beta$ hops from site $i-1$ to $i$ while no car in $\beta$ hops from $i+k-1$
 to $i+k$ under the action of ${\mathbf H}_k$. Besides, $n_k ({\mathbf H}_k
 (\beta),i) = n_k (\beta,i)-1$ if and only if a car in $\beta$ hops from site
 $i+k-1$ to $i+k$ while no car in $\beta$ hops from $i-1$ to $i$ under
 ${\mathbf H}_k$. Finally, $n_k ({\mathbf H}_k (\beta),i) = n_k (\beta,i)$ if
 and only if either no car in $\beta$ hops from sites $i-1$ and $i+k-1$ or cars
 in $\beta$ hop from both sites $i-1$ and $i+k-1$. We deduce from this
 observation together with Lemma~\ref{Lemm:Exist} that at least one site with
 $k$th order local number $n_k^{\rm (max)}$ hops under ${\mathbf H}_k$.
 Consequently, the number of cars with maximum value of $k$th order local
 number does not increase with time. In other words,
\begin{equation}
 M_k ({\mathbf H}_k (\beta),n_k^{\rm (max)} (\beta)) \leq M_k
 (\beta,n_k^{\rm (max)} (\beta)) \label{E:M_k_decrease}
\end{equation}
 and the equality holds if and only if the
 number of sites $i$ with $n_k (\beta,i) = n_k^{\rm (max)} (\beta)$ and $n_k
 ({\mathbf H}_k (\beta),i) = n_k^{\rm (max)} (\beta)-1$ equals the number of
 sites $j$ with $n_k (\beta,j) = n_k^{\rm (max)} (\beta)-1$ and $n_k
 ({\mathbf H}_k (\beta),j) = n_k^{\rm (max)}(\beta)$.
 Eq.~(\ref{E:TrafficRule_CA}) demands that $n_k (\beta,j-k)
 = n_k^{\rm (max)} (\beta)$ and hence no car can hop to the site $j-k$,
 so we also have $n_k
 ({\mathbf H}_k (\beta),j-k) = n_k^{\rm (max)} (\beta)-1$. Therefore, the
 strict inequality holds in Eq.~(\ref{E:M_k_decrease}) if there exits a site
 $i$ such that $n_k (\beta,i-k+1) = n_k^{\rm (max)} (\beta)$, $n_k (\beta,i+1)
 < n_k^{\rm (max)} (\beta)-1$, $\beta (i) = 1$ and $\beta (i+1) = 0$.

 Since we assume that $\Delta_k (\alpha) \geq 2$, we claim that we can always
 find $1\leq i\leq N$ and $j\geq 0$ such that $n_k^{\rm (max)} (\beta) = n_k
 (\beta,i-k+1) > n_k^{\rm (max)} (\beta)-1 = n_k (\beta,i+1) = n_k
 (\beta,i+k+1) = \cdots = n_k (\beta,i+jk+1) > n_k (\beta,i+(j+1)k+1)$.
 Otherwise, suppose that for some $i''$, $n_k(\beta, i''+mk+1)$ is equal to
 either $n_k^{\rm (max)} (\beta)$ or $n_k^{\rm (max)} (\beta)-1$ for all $m\in
 {\Bbb Z}$. Hence, the density $\rho$ satisfies $n_k^{\rm (max)} (\beta)-1 \leq
 k\rho \leq n_k^{\rm (max)} (\beta)$. By the assumption that $\Delta_k (\alpha)
 \geq 2$, there must be some $i'$ such that $n_k (\beta,i'+1)
 \le n_k^{\rm (max)} (\beta)-2$. The conclusion on the density forces
 that for some $m\in {\Bbb Z}$, $n_k(\beta, i'+mk+1)=n_k^{\rm (max)} (\beta)$.
 That is the claim.

 Lemma~\ref{Lemm:Exist} tells us that
 a car will hop from site $i$ to $i+1$ and hence $n_k (i-k+1)$ is reduced in no
 more than $k$ time steps. That is, there exists $r$ with $1\leq r\leq k$ such
 that $n_k ({\mathbf H}_k^r (\beta),i-k+1) < n_k (\beta,i-k+1)$. Besides,
 $n_k ({\mathbf H}_k^r (\beta),i+1) = n_k^{\rm (max)} (\beta)$ if $j \geq 2$.
 Inductively, in no more than $a \equiv k \left\lceil \frac{j+1}{2}
 \right\rceil$ time steps, number of sites with local number equals
 $n_k^{\rm (max)} (\beta)$ must be reduced by at least one. In other words,
 $M_k ({\mathbf H}_k^a (\beta),n_k^{\rm (max)} (\beta)) < M_k
 (\beta,n_k^{\rm (max)} (\beta))$. In the above discussion, it is clear that a
 packet of cars with $k$th order local number $n_k^{\rm (max)} (\beta)$
 initially located around site $i$ is moving from left to right at a speed of
 at least 1 site per time step until it hops into a region with $k$th order
 local number less than $n_k^{\rm (max)} (\beta)-1$. Thus, this packet of cars
 may prevent another packet of $k$th order local number $n_k^{\rm (max)}$ cars
 in its left hand side from moving at most $k$ times. Inductively, if
 $n_k (\beta,i-rk+1) = n_k^{\rm (max)} (\beta)$, then a car will begin to hop
 from site $i-rk$ to $i-rk+1$ in no more than $rk$ time steps. After that, this
 packet of cars will move at a speed of at least 1 site per time step from left
 to right provided that $\Delta_k$ of the configuration is still greater than
 1. By CP symmetry, a similar conclusion can be drawn for sites with $k$th
 order local number equals to $n_k^{\rm (min)}$.
 
 Since a site with $k$th order local number $n_k^{\rm (max)}$ and another site
 with $k$th order local number $n_k^{\rm (min)}$ separate by at most $N-1$
 sites, so the above discussions imply that in at most $2k \left\lceil
 \frac{1}{2} \left\lceil \frac{N}{k} \right\rceil \,\right\rceil$ time steps,
 $\Delta_k$ is reduced by at least one until $\Delta_k \leq 1$. Hence, within
 $\ell \equiv 2k (k-1) \left\lceil \frac{1}{2} \left\lceil \frac{N}{k}
 \right\rceil \,\right\rceil$ time steps, $\Delta_k < 2$.
\hfill$\Box$

\bigskip
 Combining Lemmas~\ref{Lemm:UniformDensity} and~\ref{Lemm:Fluctuation}, we know
 that the maximum fluctuation amplitude in car density evolves towards the
 minimum possible value under the repeated action of ${\mathbf H}_k$. We remark
 that relaxation time estimate $\ell$ in the above Lemma is rather
 conservative. It is not difficult to reduce this estimation by a factor of
 two or more. Nevertheless, for the purpose of solving the GDCP, the present
 estimation which states that $\ell = \mbox{O} (kN)$ for $k < N$ is already
 enough.

 Our investigation so far can be summarized in the following theorem.
\begin{Thrm}
 Let $\alpha$ be a bit string of length $N$ and $\ell \equiv \ell (k) \equiv
 2k(k-2) \left\lceil \frac{1}{2} \left\lceil \frac{N}{k} \right\rceil
 \,\right\rceil$. Then the bit string $\beta \equiv {\mathbf H}_k^\ell
 (\alpha)$ has the following properties:

 (a) The $k$th order local number $n_k (\beta,i)$ is equal to either
 $\left\lfloor k\rho \right\rfloor$ or $\left\lceil k\rho \right\rceil$;

 (b) Suppose that ${\mathbf H}_k (\beta) \neq \beta$. If $\rho (\alpha) \leq
 1/2$, then $\beta$ does not contain two consecutive $1$s as its substring.
 Similarly, if $\rho (\alpha) \geq 1/2$, then $\beta$ does not contain two
 consecutive $0$s as its substring. Thus, if $\rho (\alpha) \leq 1/2$, a car
 at site $i$ in the string $\beta$ hops to the right under ${\mathbf H}_k$ if
 and only if $\left\lceil k\rho \right\rceil = n_k (i-k+1) > n_k (i+1)=
 \left\lfloor k\rho \right\rfloor$. Similarly, if $\rho \geq 1/2$, an empty
 site $i$ in the string $\beta$ hops to the left if and only if $\left\lceil
 k\rho \right\rceil = n_k (i-k) > n_k (i) = \left\lfloor k\rho \right\rfloor$.
 \label{Thrm:TrafficRule}
\end{Thrm}
\noindent
{\it Proof:} We only need to prove~(b) as~(a) is already contained in the proof
 of Lemma~\ref{Lemm:Fluctuation}.

 To prove~(b), we use part~(a) of this Theorem. It tells us that $n_k
 (\beta,i)$ equals either $\left\lfloor k\rho \right\rfloor$ or $\left\lceil
 k\rho \right\rceil$. We denote the minimum distance between two successive
 cars in the bit string $\beta$ by $d$, that is, $d = \min \{ i-j : \beta (i) =
 1, \beta (j) = 1 \mbox{~and~} n_{i-j} (\beta,i) = 1 \}$. Since $\Delta_k = 1$,
 Lemmas~\ref{Lemm:UniformDensity} and~\ref{Lemm:Fluctuation} imply that $k\rho
 \not\in {\Bbb Z}$. Hence, the maximum distance between $\left\lceil k\rho
 \right\rceil$ cars in the string $\beta$ is greater than or equals to $k+1$.
 Thus, $d \left\lceil k\rho \right\rceil \geq k+1$. For $\rho \leq 1/2$,
\begin{eqnarray}
 d & \geq & \frac{k+1}{\left\lceil k\rho \right\rceil} \nonumber \\
 & \geq & \frac{k+1}{k/2 + 1/2} \nonumber \\
 & = & 2. \label{E:dmin}
\end{eqnarray}
 As a result, we conclude that $\beta$ does not contain the substring $11$ if
 $\rho \leq 1/2$. By CP symmetry, $\alpha$ does not contain the substring $00$
 whenever $\rho \geq 1/2$.

 The remaining assertion of part~(b) follows directly from
 Eq.~(\ref{E:TrafficRule_CA}).
\hfill$\Box$

\subsection{Separation Automaton}
\label{SS:Separation}
 With the GDCP as stated, we need to distinguish the two cases, namely,
 (1)~$\alpha=0^N$ and $f(0)\neq 0$, and (2)~$0 < \rho(\alpha)$ and
 $f(\rho(\alpha))>0$. In order to do so, we need to tell if the bit string
 $\alpha$ is equal to $0^N$ or not. It can be done using the following
 automaton ${\mathbf S}_k$ together with ${\mathbf H}_k$. In a similar way, we
 also need to distinguish the case of $\alpha = 1^N$ and $\alpha \neq 1^N$. It
 can be done using a conjugate automaton $\overline{\mathbf S}_k$. These two
 automata ${\mathbf S}_k$ and $\overline{\mathbf S}_k$ are collectively known
 as the $k$th order separation automaton.
\begin{Rul}[$k$th order separation rule]
 We denote this CA rule by ${\mathbf S}_k$. The state of site $i$ in the next
 time step ${\mathbf S}_k [\alpha](i)$ is given by:
 \begin{mathletters}
 \begin{eqnarray}
  {\mathbf S}_k [\alpha](i) = \left\{ \begin{array}{cl}
   1 & \mbox{if~} \alpha(i-k) = 1 \mbox{~and~} n_k (i-k) \\
   & ~= 1, \\
   \alpha(i) & \mbox{otherwise}.
  \end{array} \right. \label{E:SeparationRule_CA}
 \end{eqnarray}
 We denote its conjugate CA rule by $\overline{\mathbf S}_k$. That is,
 \begin{eqnarray}
  \overline{\mathbf S}_k [\alpha](i) = \left\{ \begin{array}{cl}
   0 & \mbox{if~} \alpha(i) = 0 \mbox{~and~} n_k (i-k) = \\
   & ~k-1, \\
   \alpha(i) & \mbox{otherwise}.
  \end{array} \right. \label{E:SeparationBarRule_CA}
 \end{eqnarray}
 \end{mathletters}
 \label{Rul:Separation}
\end{Rul}
\begin{Thrm}
 Let $\alpha$ be a bit string of length $N$. Then
\begin{mathletters}
\begin{equation}
 \left\{ \begin{array}{ll}
  \rho(\beta) = 0 & \mbox{if~} \rho (\alpha) = 0, \vspace{1em} \\
  \frac{1}{k} \leq \rho (\beta) < \frac{2}{k} & \mbox{if~} 0 < \rho (\alpha) <
   \frac{1}{k}, \vspace{1em} \\
  \beta = \gamma \mbox{~and~}  \rho (\beta) = \rho (\alpha) &
   \mbox{otherwise},
 \end{array} \right. \label{E:Separation_Flow}
\end{equation}
 where $\beta \equiv {\mathbf S}_k^{\left\lfloor N/k \right\rfloor} \circ
 {\mathbf H}_k^\ell (\alpha)$ and $\gamma = {\mathbf H}_k^\ell (\alpha)$.
 Similarly,
\begin{equation}
 \left\{ \begin{array}{ll}
  \rho(\delta) = 1 & \mbox{if~} \rho (\alpha) = 1, \vspace{1em} \\
  \frac{k-2}{k} \leq \rho (\delta) < \frac{k-1}{k} & \mbox{if~} \frac{k-1}{k} <
   \rho (\alpha) < 1, \vspace{1em} \\
  \delta = \gamma \mbox{~and~} \rho (\delta) = \rho (\alpha) &
   \mbox{otherwise},
 \end{array} \right. \label{E:BarSeparation_Flow}
\end{equation}
\end{mathletters}
 where $\delta \equiv \overline{\mathbf S}_k^{\left\lfloor N/k \right\rfloor}
 \circ {\mathbf H}_k^\ell (\alpha)$. \label{Thrm:SeparationRule}
\end{Thrm}
\noindent
{\it Proof:} If $\rho (\alpha) = 0$, then $\alpha = 0^N$ (this notation denotes
 $N$ consecutive $0$s and we shall use similar notations for a bit string) and
 hence $\beta = 0^N$ and $\rho (\beta) = 0$.

 If $\rho (\alpha) \geq 1/k$, then Theorem~\ref{Thrm:TrafficRule} tells us that
 $n_k (\gamma,i) > 0$ for all $i$. Hence, from Eq.~(\ref{E:SeparationRule_CA}),
 $\beta = \gamma$ and $\rho (\beta) = \rho (\alpha)$.

 Finally, if $0 < \rho (\alpha) < 1/k$, then Theorem~\ref{Thrm:TrafficRule}
 implies that $n_k (\gamma,i)$ is equal to either $0$ or $1$ for all $i$.
 Moreover, at least one bit in the string $\gamma$ is $1$. Thus, it is straight
 forward to check that applying ${\mathbf S}_k^{\left\lfloor N/k
 \right\rfloor}$ to $\gamma$ makes $1 \leq n_k (\beta,i) \leq 2$ for all $i$
 and $1/k \leq \rho (\beta) < 2/k$.

 The proof for the conjugate separation rule $\overline{\mathbf S}_k$ is
 similar.
\hfill$\Box$

\bigskip
 Thus, ${\mathbf S}_k^{\left\lfloor N/k \right\rfloor} \circ
 {\mathbf H}_k^\ell$ separates the string $0^N$ from the rest in the sense that
 no string in the form ${\mathbf S}_k^{\left\lfloor N/k \right\rfloor} \circ
 {\mathbf H}_k^\ell (\alpha)$ has a density of $1$s in the interval $(0,1/k)$.
 This is the reason why we call ${\mathbf S}_k$ the $k$th order separation
 automaton.

\subsection{Inversion Automaton}
\label{SS:Inversion}
 This CA rule and the next one (the exchange automaton) are two technical
 CA rules to transform the bit string to the form we desired. They are needed
 to correctly solve the GDCP for input strings $0^N$ and $1^N$.
\begin{Rul}[$k$th order inversion rule]
 We denote this CA rule by ${\mathbf I}_k$. The state of site $i$ in the next
 time step ${\mathbf I}_k [\alpha](i)$ is given by:
 \begin{mathletters}
 \begin{eqnarray}
  {\mathbf I}_k [\alpha](i) = \left\{ \begin{array}{cl}
   1 & \mbox{if~} n_k (i) = 0, \\
   \alpha(i) & \mbox{otherwise}.
  \end{array} \right. \label{E:InversionRule_CA}
 \end{eqnarray}
 We denote its conjugate CA rule by $\overline{\mathbf I}_k$. In other words,
 \begin{eqnarray}
  \overline{\mathbf I}_k [\alpha](i) = \left\{ \begin{array}{cl}
   0 & \mbox{if~} n_k(i) = k, \\
   \alpha(i) & \mbox{otherwise}.
  \end{array} \right. \label{E:InversionBarRule_CA}
 \end{eqnarray}
 \end{mathletters}
 \label{Rul:Inversion}
\end{Rul}
 A direct consequence of Theorem~\ref{Thrm:TrafficRule} and
 Eqs.~(\ref{E:InversionRule_CA})--(\ref{E:InversionBarRule_CA}) is:
\begin{Thrm}
 Let $\alpha$ be a bit string of length $N$ and $\beta \equiv {\mathbf I}_k
 \circ {\mathbf H}_k^\ell (\alpha)$. Then $\rho (\beta) = 1$ if $\rho (\alpha)
 = 0$, and $\rho (\beta) = \rho (\alpha)$ if $\rho (\alpha) \geq 1/k$. In fact,
 the bit string ${\mathbf H}_k^\ell (\alpha)$ is a fixed point of
 ${\mathbf I}_k$ provided that $\rho (\alpha) \geq 1/k$. Similarly, if $\gamma
 \equiv \overline{\mathbf I}_k \circ {\mathbf H}_k^\ell (\alpha)$, then $\rho
 (\gamma) = 0$ if $\rho (\alpha) = 1$, and ${\mathbf H}_k^\ell (\alpha)$ is a
 fixed point of $\overline{\mathbf I}_k$ if $\rho (\alpha) \leq (k-1)/k$.
 \label{Thrm:InversionRule}
\end{Thrm}

\subsection{Exchange Automaton}
\label{SS:Exchange}
\begin{Rul}[$k$th order exchange rule]
 We denote this CA rule by ${\mathbf E}_k$. The state of site $i$ in the next
 time step ${\mathbf E}_k [\alpha](i)$ is given by:
 \begin{eqnarray}
  {\mathbf E}_k [\alpha](i) = \left\{ \begin{array}{cl}
   1 & \mbox{if~} n_k (i) = 0, \\
   0 & \mbox{if~} n_k (i) = k, \\
   \alpha (i) & \mbox{otherwise}.
  \end{array} \right. \label{E:ExchangeRule_CA}
 \end{eqnarray}
 \label{Rul:Exchange}
\end{Rul}
 The following theorem is a direct consequence of
 Theorem~\ref{Thrm:TrafficRule} and Eq.~(\ref{E:ExchangeRule_CA}):
\begin{Thrm}
 Let $\alpha$ be a bit string of length $N$. Then $\beta \equiv {\mathbf E}_k
 \circ {\mathbf H}_k^\ell (\alpha) = 1^N$ if $\alpha = 0^N$, $\beta = 0^N$ if
 $\alpha = 1^N$. Moreover, ${\mathbf H}_k^\ell (\alpha)$ is a fixed point of
 ${\mathbf E}_k$ if $1/k \leq \rho \leq (k-1)/k$. \label{Thrm:ExchangeRule}
\end{Thrm}

\subsection{Function Automaton}
\label{SS:SmoothFunction}
 The two automata introduced in this subsection are derived from the
 continuous function in question. What we want it to achieve in the first
 automaton ${\mathbf F}_f$ is that after applying it once, there will be a
 substring of $0^{2q}$ or $1^{2q}$ according to the $\mbox{sgn}(f(\rho))$
 provided that $\rho$ is not in the form $r/q$ for some $r\in {\Bbb Z}$. This
 substring will be further manipulated by the next automaton. The second
 automaton $\tilde{\mathbf F}_f$ is used to classify a bit string with $\rho$
 in the form $r/q$ for some $r\in {\Bbb Z}$.
\begin{Rul}[function rule]
 Let $f(x)$ be the continuous function specified in GDCP. We denote the
 discontinuous points in $[0,1]$ by $p_i/q$, where $p_i,q\in {\Bbb Z}^{+} \cup
 \{ 0 \}$. (Here we choose $q$ to be the smallest positive integral
 denominators of all the expressions $p_i/q$.) Recall in
 Section~\ref{S:Problem} that we require $f(\frac{1}{2q})$ and
 $f(\frac{2q-1}{2q})$ to be non-zero. We denote the function CA rule associated
 with $f(x)$ by ${\mathbf F}_f$. The state of the site $i$ in the next time
 step ${\mathbf F}_f [\alpha](i)$ is given by:
 \begin{mathletters}
 \begin{eqnarray}
  {\mathbf F}_f [\alpha](i) = \left\{ \begin{array}{cl}
   \theta (f(\frac{2k+1}{4q})) & \mbox{if~} n_{4q} (i-j) = 2k+1 \\
   & \mbox{~for~some~} 0 \leq j \leq 2q-1, \\
   & ~k\in {\Bbb Z} \mbox{~and~} f(\frac{2k+1}{4q}) \neq 0, \\
   \alpha(i) & \mbox{otherwise},
  \end{array} \right. \label{E:FunctionRule_CA}
 \end{eqnarray}
 where
 \begin{equation}
  \theta (x) = \left\{ \begin{array}{ll}
   1 & \mbox{if~} x > 0, \\
   0 & \mbox{if~} x \leq 0.
  \end{array} \right. \label{E:theta_def}
 \end{equation}

 In addition, we denote the associated function CA rule of $f(x)$ by
 $\tilde{\mathbf F}_f$. The state of site $i$ in the next time step is given
 by:
 \begin{eqnarray}
  \tilde{\mathbf F}_f [\alpha](i) = \left\{ \begin{array}{cl}
   \theta (f(\frac{k}{2q})) & \mbox{if~} n_{2q} (i) = k, 0 < k < 2q, \\
   & ~\mbox{and~} f(\frac{k}{2q}) \neq 0, \\
  \alpha(i) & \mbox{otherwise}.
  \end{array} \right. \label{E:TildeFunctionRule_CA}
 \end{eqnarray}
 \end{mathletters}
 \label{Rul:Function}
\end{Rul}
\begin{Thrm}
 We use the notations in the function rule. Let $\alpha$ be a bit string of
 length $N$. Then $\beta \equiv {\mathbf F}_f \circ {\mathbf H}_{2q}^\ell
 (\alpha)$ equals $0^N$ if $\alpha = 0^N$, $\beta = 1^N$ if $\alpha = 1^N$.
 More importantly, if $1/2q \leq \rho (\alpha) \leq (2q-1)/2q$ and $\rho
 (\alpha)$ is not in the form $r/2q$ for some integer $r$, then $\beta$
 contains the substring $0^{2q}$ if and only if $f(\rho (\alpha)) < 0$ while
 $\beta$ contains the substring $1^{2q}$ if and only if $f(\rho(\alpha)) > 0$.
 Furthermore, ${\mathbf H}_{2q}^\ell (\alpha)$ is a fixed point of
 ${\mathbf F}_f$ provided that either $f(\rho(\alpha)) = 0$ or $\rho (\alpha)$
 is in the form $r/2q$ for some $r = 1,2,\ldots ,2q-1$.
 \label{Thrm:FunctionRule}
\end{Thrm}
\noindent
{\it Proof:} Clearly $0^N$ and $1^N$ are fixed points of both
 ${\mathbf H}_{2q}$ and ${\mathbf F}_f$. Moreover,
 Theorem~\ref{Thrm:TrafficRule} demands that $n_{2q} ({\mathbf H}_{2q}^\ell
 (\alpha),i) = r$ for all $i$ provided that $\rho (\alpha) = r/2q$ for some
 $r=1,2,\ldots ,2q-1$. Thus Eq.~(\ref{E:TrafficRule_CA}) implies in this case
 that ${\mathbf H}_{2q}^\ell (\alpha)$ is a fixed point of ${\mathbf F}_f$.
 Similarly, Eq.~(\ref{E:FunctionRule_CA}) implies that ${\mathbf H}_{2q}^\ell
 (\alpha)$ is a fixed point of ${\mathbf F}_f$ provided that $f(\rho(\alpha)) =
 0$. So it remains for us to show the validity of this Theorem when $1/2q \leq
 \rho (\alpha) \leq (2q-1)/2q$ and $\rho (\alpha)$ is not in the form $r/2q$
 for some integer $r$.

 Suppose $\rho (\alpha)$ is not in the form $r/2q$ for some integer $r$ and
 $1/2q \leq \rho (\alpha) \leq (2q-1)/2q$. Theorem~\ref{Thrm:TrafficRule}
 together with the fact that periodic boundary conditions are imposed on the
 string $\alpha$ imply that we can always find $i$ and $j$ such that $n_{2q}
 ({\mathbf H}_{2q}^\ell (\alpha),i) = n_{2q} ({\mathbf H}_{2q}^\ell
 (\alpha),i+2q)+1$ and $n_{2q} ({\mathbf H}_{2q}^\ell (\alpha),j) = n_{2q}
 ({\mathbf H}_{2q}^\ell (\alpha),j+2q)-1$. Hence, $n_{4q}
 ({\mathbf H}_{2q}^\ell (\alpha),i)$ and $n_{4q} ({\mathbf H}_{2q}^\ell
 (\alpha),j)$ are odd numbers. Since all the discontinuous points of
 $\mbox{sgn}(f)$ in the interval $[0,1]$ are in the form $p_i/q$, so for each
 fixed $r=0,1,\ldots ,2q-1$, $\mbox{sgn} (f(x)) = \mbox{sgn} (f(y))$ for all
 $x,y \in (r/2q,(r+1)/2q)$. Therefore, from Eq.~(\ref{E:FunctionRule_CA}),
 either $n_{2q} ({\mathbf F}_f \circ {\mathbf H}_{2q}^\ell (\alpha),i)$ or
 $n_{2q} ({\mathbf F}_f \circ {\mathbf H}_{2q}^\ell (\alpha),j)$ equals $2q$ if
 $f (\rho(\alpha)) > 0$. Besides, either $n_{2q} ({\mathbf F}_f \circ
 {\mathbf H}_{2q}^\ell (\alpha),i)$ or $n_{2q} ({\mathbf F}_f \circ
 {\mathbf H}_{2q}^\ell (\alpha),j)$ equals $0$ if $f (\rho(\alpha)) < 0$. As
 $1/2q \leq \rho (\alpha) \leq (2q-1)/2q$, Theorem~\ref{Thrm:TrafficRule} tells
 us that $1 \leq \left\lfloor 2q\rho (\alpha) \right\rfloor \leq n_{2q}
 ({\mathbf H}_{2q}^\ell (\alpha),i) \leq \left\lceil 2q\rho (\alpha)
 \right\rceil \leq 2q-1$ for all $i$. Hence Eq.~(\ref{E:FunctionRule_CA})
 implies that it is not possible for the string $\beta$ to contain both
 $0^{2q}$ and $1^{2q}$ as substrings. That is to say, the function rule maps a
 portion of ${\mathbf H}_{2q}^\ell (\alpha)$ to $1^{2q}$ if and only if
 $f(\rho(\alpha)) > 0$ and similarly it maps a portion of
 ${\mathbf H}_{2q}^\ell (\alpha)$ to $0^{2q}$ if and only if $f(\rho(\alpha)) <
 0$.
\hfill$\Box$

\begin{Thrm}
 We use the notations of the function rule. Let $\alpha$ be a bit string of
 length $N$. Then $\beta \equiv \tilde{\mathbf F}_f \circ {\mathbf H}_{2q}^\ell
 (\alpha)$ equals $0^N$ if $\rho (\alpha) = r/2q$ for some $r = 1,2,\ldots
 ,2q-1$ and $f(r/2q) < 0$. Similarly, $\beta = 1^N$ if $\rho (\alpha) = r/2q$
 for some $r=1,2,\ldots ,2q-1$ and $f(r/2q) > 0$. Moreover,
 ${\mathbf H}_{2q}^\ell (\alpha)$ is a fixed point of $\tilde{\mathbf F}_f$ if
 (a)~$\alpha = 0^N$, or (b)~$\alpha = 1^N$, or (c)~$0 < \rho (\alpha) < 1$ and
 $f(\rho (\alpha)) = 0$. \label{Thrm:TildeFunctionRule}
\end{Thrm}
\noindent
{\it Proof:} Direct application of Theorem~\ref{Thrm:TrafficRule} and
 Eq.~(\ref{E:TildeFunctionRule_CA}).
\hfill$\Box$

\subsection{Propagation Automaton}
\label{SS:Propagation}
 After applying the function automaton ${\mathbf F}_f$, there will be a
 substring of $0^{2q}$ or $1^{2q}$ provided that $\rho$ is not in the form
 $r/q$ for some $r\in {\Bbb Z}$ and $f(\rho)\neq 0$. To convert the whole bit
 string to the desired form, we have to propagate the substring to everywhere.
\begin{Rul}[propagation rule]
 Let $\alpha$ be a bit string. We
 denote the propagation rule by ${\mathbf P}_f$. The effect of ${\mathbf P}_f$
 on each bit is given by
 \begin{eqnarray}
  {\mathbf P}_f [\alpha](i) = \left\{ \begin{array}{cl}
   1 & \mbox{if~} n_{2q} (i+1) = 2q \\
   0 & \mbox{if~} n_{2q} (i+1) = 0, \\
   \alpha (i) & \mbox{otherwise}.
  \end{array} \right. \label{E:PropagationRule_CA}
 \end{eqnarray}
 \label{Rul:Propagation}
\end{Rul}
\begin{Thrm}
 We use the notations of the function rule. Let $\alpha$ be a bit string of
 length $N$ and $\beta \equiv \tilde{\mathbf F}_f \circ {\mathbf P}_f^{N-2q}
 \circ {\mathbf F}_f \circ {\mathbf H}_{2q}^{\ell (2q)} (\alpha)$.

 (a)~The strings $0^N$ and $1^N$ are fixed points of $\tilde{\mathbf F}_f \circ
 {\mathbf P}_f^{N-2q} \circ {\mathbf F}_f \circ {\mathbf H}_{2q}^{\ell (2q)}$.
 Moreover, $\beta = 0^N$ provided that $0 < \rho (\alpha) < 1/2q$ and
 $f(\frac{1}{2q}) < 0$. Similarly, $\beta = 1^N$ provided that $(2q-1)/2q <
 \rho (\alpha) < 1$ and $f(\frac{2q-1}{2q}) > 0$.

 (b)~Suppose $1/2q \leq \rho (\alpha) \leq (2q-1)/2q$. Then $\beta = 0^N$ if
 and only if $f(\rho(\alpha)) < 0$. Moreover, $\beta = 1^N$ if and only if
 $f(\rho(\alpha)) > 0$. Finally, $\beta = {\mathbf H}_{2q}^\ell (\alpha)$ and
 hence $\rho (\beta) = \rho(\alpha)$ if and only if $f(\rho(\alpha)) = 0$.
 \label{Thrm:PropagationRule}
\end{Thrm}
\noindent
{\it Proof:} To show that $0^N$ and $1^N$ are fixed points of
 $\tilde{\mathbf F}_f \circ {\mathbf P}_f^{N-2q} \circ {\mathbf F}_f \circ
 {\mathbf H}_{2q}^\ell$ is straight forward. Suppose $0 < \rho (\alpha) <
 1/2q$ and $f(1/2q) < 0$. Then, Theorem~\ref{Thrm:TrafficRule} implies that
 $n_{2q} ({\mathbf H}_{2q}^\ell (\alpha),i) \leq 1$ with equality holds for
 some $i$. Since $f(1/2q) < 0$, so by continuity of $f$ and the distribution of
 discontinuous points of $\mbox{sgn} (f)$, we know that $f(x) < 0$ whenever $0
 < x < 1/q$. Hence ${\mathbf F}_f \circ {\mathbf H}_{2q}^\ell (\alpha)$
 contains the substring $0^{2q}$ and does not contain the substring $1^{2q}$.
 Thus by Eq.~(\ref{E:PropagationRule_CA}) and
 Theorem~\ref{Thrm:TildeFunctionRule}, we conclude that $\beta = 0^N$. The
 proof of $\beta = 1^N$ provided that $(2q-1)/2q < \rho (\alpha) < 1$ and
 $f(\frac{2q-1}{2q}) > 0$ is similar. So, we have proved the validity of
 part~(a).

 We know from Theorem~\ref{Thrm:TrafficRule} that
 ${\mathbf H}_{2q}^\ell (\alpha)$ is a fixed point of ${\mathbf P}_f$ provided
 that $1/2q \leq \rho (\alpha) \leq (2q-1)/2q$. Therefore, part~(b) of this
 Theorem follows directly from Theorems~\ref{Thrm:FunctionRule},
 \ref{Thrm:TildeFunctionRule} and Eq.~(\ref{E:PropagationRule_CA}).
\hfill$\Box$

\section{Generalized Density Classification Automata}
\label{S:Solution}
 After introducing all the necessary building blocks in the last section, we
 are ready to report our solution of the GDCP. 
\begin{Thrm}
 Let $\alpha$ be a bit string of length $N$. Let $f(x)$ be a continuous
 function. Denote the rational discontinuous points of $\mbox{sgn} (f(x))$ by
 $p_i/q$ with $p_i, q \in {\Bbb Z}^{+} \cup \{ 0 \}$. (Here $q$ is chosen to be
 the smallest positive integral denominator for the expressions $p_i/q$.)
 Suppose further that $f(\frac{1}{2q})$ and $f(\frac{2q-1}{2q})$ are both
 non-zero, then the following sequence of CA rules solves the GDCP for the
 function $f(x)$:
 \begin{eqnarray}
  {\mathbf C}_f & = & \tilde{\mathbf F}_f \circ {\mathbf P}_f^{N-2q} \circ
  {\mathbf F}_f \circ \left( {\mathbf E}_{2q} \right)^{a5} \circ \left(
  \overline{\mathbf I}_{2q} \right)^{a4} \circ \left( {\mathbf I}_{2q}
  \right)^{a3} \circ \nonumber \\
  & & \left( {\mathbf H}_{2q}^{\ell (2q)} \circ
  \overline{\mathbf S}_{2q}^{\left\lfloor N/2q \right\rfloor} \right)^{a2}
  \circ \left( {\mathbf H}_{2q}^{\ell (2q)} \circ
  {\mathbf S}_{2q}^{\left\lfloor N/2q \right\rfloor} \right)^{a1} \circ
  \nonumber \\
  & & {\mathbf H}_{2q}^{\ell (2q)}, \label{E:Method}
 \end{eqnarray}
 where $a1 = 1$ if $f(\frac{1}{2q}) > 0$, $a1 = 0$ otherwise; $a2 = 1$ if
 $f(\frac{2q-1}{2q}) < 0$, $a2 = 0$ otherwise; $a3 = 1$ if $f(0) > 0$ and $f(1)
 \geq 0$, $a3 = 0$ otherwise; $a4 = 1$ if $f(1) < 0$ and $f(0) \leq 0$, $a4 =
 0$ otherwise; $a5 = 1$ if $f(0) > 0$ and $f(1) < 0$ and $a5 = 0$ otherwise.
 \label{Thrm:Main}
\end{Thrm}

 Before we prove this theorem, let us first informally explain how it works.
 The ${\mathbf H}_{2q}^{\ell (2q)}$ in the right-most part of
 Eq.~(\ref{E:Method}) makes the density of car uniform. Then the
 $\left( {\mathbf H}_{2q}^{\ell (2q)} \circ {\mathbf S}_{2q}^{\left\lfloor N/2q
 \right\rfloor} \right)^{a1}$ term first separates $\rho = 0$ from $0 < \rho <
 1/2q$ and then makes the resultant bit string uniform in car density whenever
 $f(\frac{1}{2q})$ and hence also $f(x) > 0$ for all $0 < x < 1/2q$. Similarly,
 the $\left( {\mathbf H}_{2q}^{\ell (2q)} \circ
 \overline{\mathbf S}_{2q}^{\left\lfloor N/2q \right\rfloor} \right)^{a2}$ term
 first separates $\rho = 1$ from $(2q-1)/2q < \rho < 1$ and then makes the
 resultant bit string uniform in car density whenever $f(\frac{2q-1}{2q})$ and
 hence also $f(x) < 0$ for all $(2q-1)/2q < x < 1$. Next, 
 notice that $a3$ and $a5$ will not simultaneously equal to one,
 and similarly $a4$ and $a5$ will not simultaneously equal to one, the $\left(
 {\mathbf E}_{2q} \right)^{a5} \circ \left( \overline{\mathbf I}_{2q}
 \right)^{a4} \circ \left( {\mathbf I}_{2q} \right)^{a3}$ term correctly deals
 with the GDCP for $\rho = 0$ and $1$. This is followed by the term
 ${\mathbf P}_f^{N-2q} \circ {\mathbf F}_f$ which correctly classifies those
 $\rho$ not in the form $r/2q$. Finally, the term $\tilde{\mathbf F}_f$ settles
 the remaining case of $\rho$ in the form $r/2q$ with $r\neq 0$ or $2q$.

\bigskip\noindent
{\it Proof:} We divide the proof into the following four cases:

 Case~(a): $\rho (\alpha) = 0$ or $1$, that is, $\alpha = 0^N$ or $1^N$: in
 this case, ${\mathbf C}_f = \tilde{\mathbf F}_f \circ {\mathbf P}_f^{N-2q}
 \circ {\mathbf F}_f \circ \left( {\mathbf E}_{2q} \right)^{a5} \circ \left(
 \overline{\mathbf I}_{2q} \right)^{a4} \circ \left( {\mathbf I}_{2q}
 \right)^{a3}$. So by Theorems~\ref{Thrm:InversionRule}
 and~\ref{Thrm:ExchangeRule}, ${\mathbf C}_f (0^N) = \tilde{\mathbf F}_f \circ
 {\mathbf P}_f^{N-2q} \circ {\mathbf F}_f (x^N)$ where $x = \mbox{sgn} (f(0))$
 and similarly ${\mathbf C}_f (1^N) = \tilde{\mathbf F}_f \circ
 {\mathbf P}_f^{N-2q} \circ {\mathbf F}_f (y^N)$ where $y = \mbox{sgn} (f(1))$.
 Hence, from Eqs.~(\ref{E:FunctionRule_CA}), (\ref{E:TildeFunctionRule_CA})
 and~(\ref{E:PropagationRule_CA}), the theorem holds for $\rho (\alpha) = 0$ or
 $1$.

 Case~(b): $1/2q \leq \rho (\alpha) \leq (2q-1)/2q$: in this case,
 Theorems~\ref{Thrm:SeparationRule}, \ref{Thrm:InversionRule}
 and~\ref{Thrm:ExchangeRule} tell us that ${\mathbf C}_f = \tilde{\mathbf F}_f
 \circ {\mathbf P}_f^{N-2q} \circ {\mathbf F}_f \circ {\mathbf H}_{2q}^{\ell
 (2q)}$. Hence, this case is settled by applying
 Theorem~\ref{Thrm:PropagationRule}.

 Case~(c): $0 < \rho (\alpha) < 1/2q$: in this case, the continuity of $f$
 together with our assumptions that $f(1/2q) \neq 0$ and discontinuous points
 of $\mbox{sgn} (f(x))$ are in the form $r/q$ for some $r\in {\Bbb Z}$ demand
 that $f$ can be further divided into the following two subcases, namely,
 (1)~$f(x) > 0$ for all $0 < x < 1/q$, and (2)~$f(x) < 0$ for all $0 < x <
 1/q$.

 In subcase~(1), $a1 = 1$. Theorem~\ref{Thrm:SeparationRule} tells us that
 $1/2q \leq \rho ({\mathbf S}_{2q}^{\left\lfloor N/2q \right\rfloor} \circ
 {\mathbf H}_{2q}^\ell (\alpha)) < 1/q$. From Theorem~\ref{Thrm:InversionRule},
 we know that ${\mathbf C}_f (\alpha) =
 \tilde{\mathbf F}_f \circ {\mathbf P}_f^{N-2q} \circ {\mathbf F}_f \circ
 {\mathbf H}_{2q}^{\ell (2q)} \circ {\mathbf S}_{2q}^{\left\lfloor N/2q
 \right\rfloor} \circ {\mathbf H}_{2q}^{\ell (2q)} (\alpha)$. Combining with
 Theorem~\ref{Thrm:PropagationRule}, ${\mathbf C}_f (\alpha) = 1^N =
 (\mbox{sgn} (f(\rho)))^N$. 

 In subcase~(2), the continuity of $f$ implies that $a1 = a3 = a5 = 0$ and
 hence ${\mathbf C}_f (\alpha) = \tilde{\mathbf F}_f \circ {\mathbf P}_f^{N-2q}
 \circ {\mathbf F}_f \circ {\mathbf H}_{2q}^{\ell (2q)} (\alpha)$. So, applying
 Theorem~\ref{Thrm:PropagationRule} gives ${\mathbf C}_f (\alpha) = 0^N =
 (\mbox{sgn} (f(\rho)))^N$.

 Therefore, Eq.~(\ref{E:Method}) correctly classifies bit string with $0 < \rho
 < 1/2q$.

 Case~(d): $(2q-1)/2q < \rho (\alpha) < 1$: The proof of this case is similar
 to that of case~(c) and we are not going to write the details here.

 In summary, ${\mathbf C}_f (\alpha) = 0^N$ if $f(\rho) < 0$, ${\mathbf C}_f
 (\alpha) = 1^N$ if $f(\rho) > 0$, and ${\mathbf C}_f (\alpha) =
 {\mathbf H}_{2q}^{\ell (2q)} (\alpha)$ if $f(\rho) = 0$. Consequently, the
 result of the GDCP can be read out from any $2q$ consecutive sites of the
 final output bit string. More precisely, if we find that such substring equals
 $0^{2q}$, then either $f(\rho) < 0$ or $\rho = 0$ and $f(0) = 0$. If such
 substring is $1^{2q}$, then either $f(\rho) > 0$ or $\rho = 1$ and $f(1) = 0$.
 Lastly, if such substring contains both 0 and 1, then $\rho \neq 0$, $\rho
 \neq 1$ and $f(\rho) = 0$.
\hfill$\Box$

\section{Simple Solutions To Certain Special Cases}
\label{S:Generalization}
 Theorem~\ref{Thrm:Main} provides a solution to the GDCP involving all the nine
 automata introduced in Section~\ref{S:BuildingBlock_CAs}. Nonetheless, it does
 not mean that solution of \emph{any} GDCP have to be that complicated. In this
 section, we report simple solutions to certain useful GDCPs. Comparing to the
 GDCP, solutions to the following problems are rather straight-forward. So the
 presentation in this section is brief.
\subsection{Cases Related To Rational Density Classification}
\label{SS:RationalDC}
 The simplest non-trivial case of GDCP is the rational density problem which
 chooses $f(x) = x - \rho_c$ for a fixed rational number $0 < \rho_c < 1$. Chau
 \emph{et al.} solved this problem by the following two automata
 \cite{RationalDC}.
\begin{Rul}[modified traffic rule]
 Let $\alpha$ be a bit string of length $N$ and $\rho_c = p/q$ where $p,q \in
 {\Bbb Z}^{+}$ with $p$ and $q$ are relatively prime. Then the modified
 traffic rule is give by
 \begin{equation}
  {\mathbf T}_\rho [\alpha] (i) = \left\{ \begin{array}{cl}
   1 & \mbox{if~} \alpha(i) = 0, \alpha(i-1) = 1, \mbox{~and} \\
   & ~n_{q-1} (i) \leq p-1, \\
   0 & \mbox{if~} \alpha(i) = 1, \alpha(i+1) = 0, \mbox{~and} \\
   & ~n_{q-1} (i+1) \leq p-1, \\
   \alpha(i) & \mbox{otherwise}.
  \end{array} \right. \label{E:MTR_CA}
 \end{equation}
\end{Rul} 
\begin{Rul}[Modified Majority Rule]
 Let $\alpha$ be a bit string of length $N$ and $\rho_c = p/q$ where $p,q \in
 {\Bbb Z}^{+}$ with $p$ and $q$ are relatively prime. Then the modified
 majority rule is give by
 \begin{equation}
  {\mathbf M}_\rho [\alpha] (i) = \left\{ \begin{array}{ll}
   1 & \mbox{if~} n_{2q+1} (i-q) \geq 2p+1, \\
   0 & \mbox{otherwise}.
  \end{array} \right. \label{E:MMR_CA}
 \end{equation}
\end{Rul}
 More precisely, they showed that ${\mathbf M}^{\left\lceil N/2(q-1)
 \right\rceil} \circ {\mathbf T}^{\left\lceil N (\max (q,2p)-1) \max (q-p,p)/pq
 \right\rceil +q-2}$ solves the rational density classification problem with
 $f(x) = x-\rho_c$.

 By applying ${\mathbf I}_q$, $\overline{\mathbf I}_q$ or ${\mathbf E}_q$ once
 to the resultant state of the rational density classification automata, it is
 clear that we can solve a number of related problems using three CAs,
 including $f(x) = \rho_c - x$, $f(x) = (x-\rho_c)^2$ and $f(x) =
 -(x-\rho_c)^2$.

\subsection{Cases Related To Coarse-Grained Rational Density Classification}
\label{SS:CoarseGrainDC}
 Another interesting and useful problem we have briefly mentioned in
 Section~\ref{S:Intro} is the so-called approximate rational density
 classification. Let us recall that if $\rho_c = p/q$ is a rational number,
 then solution to the rational density classification problem either by the
 method reported in Section~\ref{S:Solution} or in Ref.~\cite{RationalDC}
 requires a rule table whose range scales linearly with $q$. Thus, it is more
 cost effective to classify a coarse-grained rational density. To achieve this
 task, we are required to solve the GDCP with the function $f$ given by
 Eq.~(\ref{E:CoarseGrainDC}) with $\rho_{c1} < \rho_c < \rho_{c2}$. An
 effective way to choose $\rho_{c1}$ and $\rho_{c2}$ is to use the continued
 fraction approximation of $\rho_c$. If the output string is $1^N$, we know
 that $\rho$ is greater than $\rho_{c2}$ and hence also $\rho_c$. Similarly,
 we know that $\rho < \rho_c$ if the output is $0^N$. If $\rho_{c1} < \rho < 
 \rho_{c2}$, then we do not know if the $\rho$ is greater than $\rho_c$ or not.
 Fortunately, $\rho$ is preserved by the automata in this case and hence we may
 feed our output bit string to say the full rational density classification
 CAs for fine-grained density determination. In this way, we can efficiently
 solve the rational density classification problem with small rule table size
 with high probability for a randomly given input bit string. Performing a CA
 which is a function of output of another CA leads us to the notion of CA
 programming. We shall explore the power and weakness of CA programming
 elsewhere \cite{CAProgramming}.

 After pointing out the significance of the coarse-grained rational density
 classification problem, we report a solution involving only two CA rules. We
 write $\rho_{c1} = p_1/q$ and $\rho_{c2} = p_2/q$ as usual. The first CA rule
 is our car hopping automaton ${\mathbf H}_q$. (Clearly we cannot use
 ${\mathbf T}$ as we have two critical densities here.) The second CA rule
 is a variation of the propagation rule, as stated below.
\begin{Rul}[modified propagation rule]
 Let $\alpha$ be a bit string. Then the modified propagation rule is given by
 \begin{equation}
  \tilde{\mathbf P}_{p_1,p_2,q} [\alpha] (i) = \left\{ \begin{array}{cl}
   1 & \mbox{if~} n_q (i+j) > p_2 \mbox{~for~some} \\
   & ~1\leq j\leq q, \\
   0 & \mbox{if~} n_q (i+1) < p_1 \mbox{~for~some} \\
   & ~1\leq j\leq q, \\
   \alpha(i) & \mbox{otherwise}.
  \end{array} \right. \label{E:TildePropagation_CA}
 \end{equation}
\end{Rul}
 Clearly, if $\rho > \rho_{c2}$, there is a site $i$ such that $n_q
 ({\mathbf H}_q^\ell (\alpha),i) > p_2$. Thus, $\rho$ increases to $1$ under
 the repeated application of $\tilde{\mathbf P}_{p_1,p_2,q}$. Similarly, $\rho$
 decreases to $0$ under the repeated application of
 $\tilde{\mathbf P}_{p_1,p_2,q}$ if $\rho < \rho_{c1}$. In summary, we know
 that ${\mathbf P}_{p_1,p_2,q}^{\left\lceil N/q \right\rceil} \circ
 {\mathbf H}_q^{\ell (q)}$ solves the coarse-grained rational density
 classification problem.

 In a similar way, problems with $f(x) = (x-\rho_{c1}) (x-\rho_{c2})$, $f(x) =
 (x-\rho_{c1})(x-\rho_{c2})^2$ and so on can be solved using four CAs.
  
\subsection{Variation Of The Theme}
\label{SS:Variation}
 Because of the difficulties in distinguishing between the strings $0^N$ and
 $0^{N-1}1$, we introduce the separation automaton ${\mathbf S}_k$.
 Unfortunately, ${\mathbf S}_k$ is not car density conserving. This is
 precisely the reason why we impose the technical conditions that $f(1/2q) \neq
 0$ and $f((2q-1)/2q) \neq 0$. We remark that the above technical conditions
 can be waived provided that we relax the GDCP a bit. Instead of requiring that
 the density of 1s of the output string $\beta$ equals to that of the input
 string $\alpha$ if $f(\rho(\alpha)) = 0$, we replace it by requiring that
 $|f(\rho(\beta)) - f(\rho(\alpha))| < 1/q$. By doing so, it is clear that the
 GDCP can be solved even when $f(1/2q)$ or $f((2q-1)/2q) = 0$ by the
 following sequence of automata:
\begin{eqnarray}
 & & \tilde{\mathbf F}_f \circ {\mathbf P}_f^{N-2q} \circ {\mathbf F}_f \circ
  \left( {\mathbf E}_{2q} \right)^{a5} \circ \left( \overline{\mathbf I}_{2q}
  \right)^{a4} \circ \left( {\mathbf I}_{2q} \right)^{a3} \circ \nonumber \\
 & & {\mathbf H}_{2q}^{\ell (2q)} \circ \overline{\mathbf S}_{2q}^{\left\lfloor
  N/2q \right\rfloor} \circ {\mathbf H}_{2q}^{\ell (2q)} \circ
  {\mathbf S}_{2q}^{\left\lfloor N/2q \right\rfloor} \circ
  {\mathbf H}_{2q}^{\ell (2q)}, \label{E:AltMethod}
\end{eqnarray}
 where $a3 = 1$ if $f(0) > 0$ and $f(1) \geq 0$, $a3 = 0$ otherwise; $a4 = 1$
 if $f(1) < 0$ and $f(0) \leq 0$, $a4 = 0$ otherwise; $a5 = 1$ if $f(0) > 0$
 and $f(1) < 0$ and $a5 = 0$ otherwise.

 However, we do not encourage the use of this relaxed definition of GDCP for
 the density of 1s is not preserved in case $f(\rho) = 0$.
\section{Discussions}
\label{S:Discussion}
 A few remarks are in order. First, Theorem~\ref{Thrm:Main} provides a CA
 solution to the GDCP with at most eight CA rules each with rule table of range
 $\leq 4q$. (Note that the number eight comes from the observation that we can
 combine ${\mathbf F}_f \circ \left( {\mathbf E}_{2q} \right)^{a5} \circ \left(
 \overline{\mathbf I}_{2q} \right)^{a4} \circ \left( {\mathbf I}_{2q}
 \right)^{a3}$ together to form one single CA rule whose rule table size is
 still independent of the string length $N$.) Moreover, the total run time
 required scales as $\mbox{O} (qN)$ whenever $q < N$, making it asymptotically
 optimal up to a constant factor. The main ingredient used to solve the GDCP is
 the $k$th order traffic rule ${\mathbf H}_k$ whose repeated application leads
 to a uniformly distributed $1$s in the bit string in the sense that
 fluctuation of the $k$th order local number $n_k (i)$ does not exceed $1$.

 We stress that the solution to the GDCP using a sequence of CAs is not unique.
 We have also found that a slightly
 different traffic rule involving $2k+1$ sites
 together with a generalized majority vote rule based on the work in
 Ref.~\cite{RationalDC} can also do the job. Nevertheless, this alternative
 method requires in general more than eight CA rules in succession
 \cite{YanSiu}.

 We are not sure if the GDCP can be solved using fewer than eight CA rules.
 What we know from the result of Fuk\'{s} is that density classification, being
 a special case of GDCP, cannot be solved using a single CA rule\cite{Fuks}.
 Although certain special cases of the GDCP, such as the original density
 classification problem can be solved using two CA rules\cite{Fuks,RationalDC},
 we feel that it is highly unlikely to solve the full GDCP using just two CA
 rules because of the difficulties involved in separating the cases with $\rho$
 close to $0$ and $\rho$ equals $0$ although we have provided solutions of
 the rational density classification and coarse-grained rational density
 classification problems using two CA rules in Section~\ref{S:Generalization}.

 Finally, we remark that the CA solution to the GDCP shows a rich flow diagram. 
 The flow of $\rho$ under the action of ${\mathbf C}_f$ exhibits at most two
 stable fixed points at $\rho = 0$ and $1$ corresponding to $f(\rho) \neq 0$,
 finitely many fixed points corresponding to the rational isolated roots of the
 equation $f(x) = 0$, together with possibly infinitely many neutral fixed
 points corresponding to the remaining non-isolated zeros of $f(x) = 0$ in the
 limit of $N\rightarrow\infty$. In this respect, sequence of CAs may lead to
 very interesting flow diagrams that are distinctive from those resulting from
 conventional continuous dynamical systems.
\acknowledgements
 This work is supported in part by the Hong Kong SAR Government RGC grant
 HKU~7098/00P. H.F.C. is also supported in part by the Outstanding Young
 Researcher Award of the University of Hong Kong.

\end{multicols}
\end{document}